\documentclass[aps,prl,letterpaper,twocolumn,nofootinbib,preprintnumbers,floatfix,letterpaper]{revtex4}
\pdfoutput=1 
\usepackage{graphicx,dcolumn,bm,epsfig,cancel,hyperref}
\usepackage{amsmath}
\usepackage{amssymb, times}
\begin{document}
\title{Supernova Cooling in a Dark Matter Smog}
\author{Yue Zhang}
\affiliation{California Institute of Technology, Pasadena, CA 91125}

\begin{abstract}
A light hidden gauge boson with kinetic mixing with the usual photon is a popular setup in theories of dark matter. The supernova cooling via radiating the hidden boson is known to put an important constraint on the mixing. I consider the possible role dark matter, which under reasonable assumptions naturally exists inside supernova, can play in the cooling picture. Because the interaction between the hidden gauge boson and DM is likely unsuppressed, even a small number of dark matter compared to protons inside the supernova could dramatically shorten the free streaming length of the hidden boson. A picture of a dark matter ``smog'' inside the supernova, which substantially relaxes the cooling constraint, is discussed in detail.
\end{abstract}

\preprint{CALT-68-2888}

\maketitle

In spite of the great triumph of Standard Model (SM) of particle physics, there are compelling reasons for going beyond it, one of which is to understand the nature of dark matter (DM) in our universe. If due to a particle physics origin, DM can be viewed to belong to a hidden sector.
A hidden sector can be complicated, containing degrees of freedom other than the DM itself.
The massive gauge boson of a hidden U(1) interaction can arise from many well motivated theories~\cite{Holdom:1986eq, Galison:1983pa, Pospelov:2007mp, ArkaniHamed:2008qn, Cheung:2009qd, An:2009vq, Foot:2010hu}.
It can play an important role in the DM phenomenology, serving as a portal from the hidden sector to the SM sector.
Therefore, hidden gauge boson is one of the candidates widely searched for at the cosmic and intensity frontiers~\cite{Essig:2013lka}.

The observation of supernova (SN) 1987a can impose a powerful constraint on the kinetic mixing between the usual photon and an MeV to GeV scale hidden gauge boson. 
Radiating too much energy to space via the hidden boson will affect the observed SN neutrino spectrum in the first few seconds.
It was shown~\cite{Bjorken:2009mm, Dent:2012mx, Dreiner:2013mua} for hidden gauge boson mass below 100\,MeV, the cooling argument has excluded a window between $10^{-7}$\,--\,$10^{-10}$ for the kinetic mixing. In together with other low energy constraints, the mixing is bounded to be less than $10^{-10}$. 
It is worth noting that, to obtain these constraints only the interactions between the hidden gauge boson and SM particles are included, but the interaction with DM has been neglected.

Since the interests in hidden gauge boson is largely motivated by the study of DM, in this {\it letter}, I consider the possible role DM can play in the cooling dynamics of SN, 
and how the SN constraints have to be reinterpreted.
I will assume the DM mass is much larger than the SN temperature so itself cannot be produced by the SN.
However, it is natural to expect DM to exist inside SN, because the progenitor of SN was a star and should capture the DM it met with throughout the lifetime.
Since the hidden gauge boson interacts with DM, the presence of DM forms a smog inside and near the core, which increases the opacity to the hidden boson.
As a result, the constraint on kinetic mixing could be weakened.
The core of SN (young neutron star) is a {\it unique} place for this effect to be significant.
As shown below, it has a relatively high temperature but relatively small volume, sufficient for the dark matter ``smog'' to fully embrace the core region.

To be specific, I consider a simple dark matter sector containing a hidden U(1) theory with kinetic mixing with the SM photon. 
The dark matter carries a unit hidden charge.
\begin{eqnarray}\label{theory}
\begin{split}
\mathcal{L}_{\rm dark} = &- \frac{\varepsilon}{2} F_{\mu\nu} F'^{\mu\nu} + m^2_{A'} A'_\mu A'^\mu \\
&+ \bar\chi i\gamma^\mu(\partial_\mu - i e' A'_\mu) \chi + M_\chi \bar\chi \chi \ ,
\end{split}
\end{eqnarray}
where $A'_\mu$ is the hidden gauge boson, or dark photon.
This Lagrangian can be obtained in a complete theory when hidden U(1) first mixes with hypercharge~\cite{Babu:1997st}.
It is useful to redefine the photon field $A_\mu\to A_\mu - \varepsilon A'_\mu$ to remove the kinetic mixing term.
In the new basis, QED remain unchanged but all the SM fermions feel the hidden U(1) gauge interaction, {\rm {\it i.e.}},
a fermion $f$ with electric charge $q_f$ also carries a hidden charge $\sim\varepsilon q_f$.

\begin{figure}[h]
\vspace{-0.2cm}
\includegraphics[width=0.9\columnwidth]{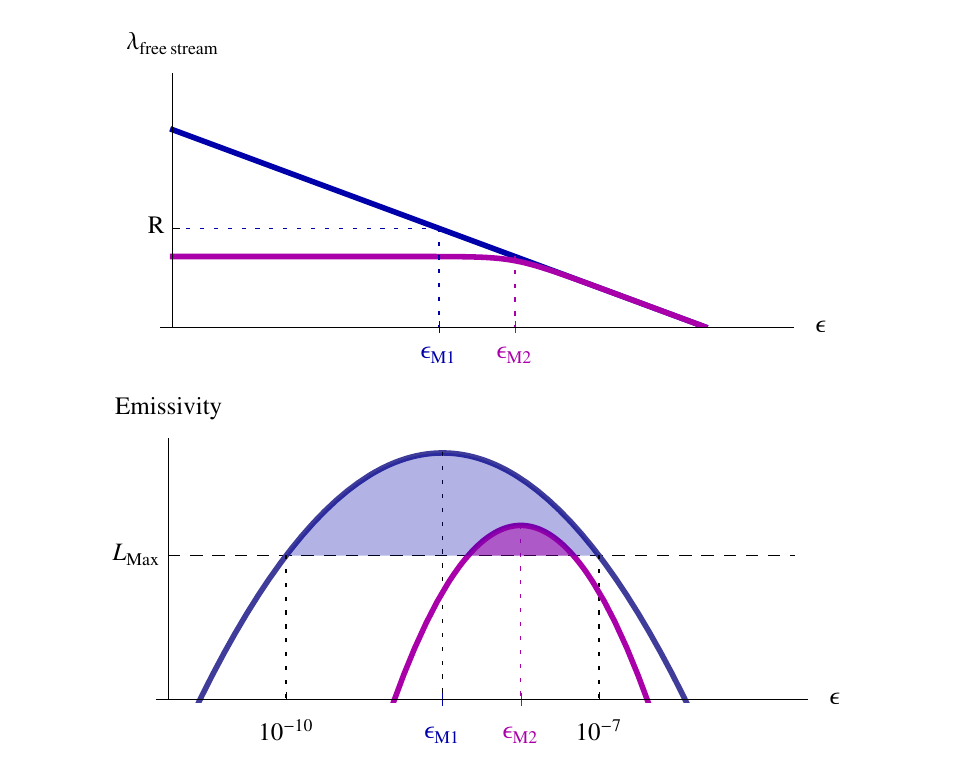}
\vspace{-0.2cm}
\caption{Schematic plots on the $\varepsilon$ dependence in $A'$'s free streaming length (upper) and the emissivity from SN (lower).
Blue (magenta) illustrates SN emitting $A'$ without (with) the presence of DM.
}\label{fig:pic}
\vspace{-0.3cm}
\end{figure}

The interaction between proton and $A'$ plays an important role in cooling the SN (see blue curves in Fig.~\ref{fig:pic}).
The relevant processes are: bremsstrahlung from proton scattering $pp\to ppA'$ which produces $A'$, and inverse bremsstrahlung $ppA'\to pp$ for the absorption. 
They dominate over the other processes such as $p\gamma\leftrightarrow p A'$ because the proton has much higher number density than other particles in the core region of SN.
Increasing $\varepsilon$ increase the production rate, but also shortens the free streaming length.
For small $\varepsilon$, the emissivity of $A'$ first increases as $\varepsilon$ grows until its streaming length reduces to the size of SN, $R$. Afterwards, an $A'$ sphere emerges inside which $A'$ is trapped and thermalized.
Emitting $A'$ from the surface of the sphere still cools the SN core, but the emissivity decreases as $\varepsilon$ grows. 

This picture could be changed dramatically if there are also DM $\chi$ inside the SN. 
The interaction of $A'$ with DM is typically much stronger than with proton, therefore, 
even very little amount of DM could significantly modify the picture of $A'$ emission.
In contrast to the proton case, the DM number density is much lower, which highly suppresses
bremsstrahlung processes $\chi\chi\leftrightarrow\chi\chi A'$.
The most important process for $A'$ to interact with DM is via Thomson scattering $A'\chi\to A'\chi$.
This means the DM is better at deflecting/trapping $A'$ inside SN than producing them.
The free streaming length in this case is,
\begin{eqnarray}
\lambda_{\rm fs} = \frac{1}{n_p \sigma_{p A'} + n_\chi \sigma_{\chi A'}} \ ,
\end{eqnarray}
where $\sigma_{p A'}$ stands for $ppA'\to pp$ cross section and $\sigma_{\chi A'}$ for $A'\chi\to A'\chi$.
As said, although the second term in the denominator is suppressed by a factor $n_\chi/n_p$, 
the cross section $\sigma_{\chi A'}$ could be much higher than $\sigma_{p A'}$
due to the lack of $\varepsilon^2$ suppression. When $\varepsilon^2 \ll n_\chi/n_p$, the ${\chi A'}$ term dominates.

This leads to an interesting possibility when $n_p \sigma_{p A'}\ll R^{-1} < n_\chi \sigma_{\chi A'}$.
In this case, $\lambda_{\rm fs}$ is smaller than $R$ --- the Thomson scattering with DM creates a {\it pseudo} $A'$ sphere. 
I call it a pseudo sphere because $A'$ cannot escape from its inside, however, the production rate (due to $\sigma_{p A'}$) is not large enough to thermalize $A'$.
In other words, $A'$ is in kinetic equilibrium in the pseudo sphere, but not yet in chemical equilibrium --- the presence of DM simply functions as a ``smog'' to $A'$.
Then the cooling is dominated by emission from either the rest of the core outside the sphere, or its surface but at a suppressed rate than the black-body radiation (see magenta curves in Fig.~\ref{fig:pic}).
This offers an opportunity to reopen part of the excluded window of $\varepsilon$. 

The rest of this paper aims at making more quantitative statement on the above central point, and classifying the phases of the cooling process.

To set the stage, I start with the conventional picture with no DM inside the SN.
In the first seconds, the SN 1987a can be modeled~\cite{Burrows:1986me, Mayle:1986ic, Turner:1987by, Barbieri:1988av} by a core with constant temperature $T_c = 30\,$MeV, constant number density of protons $n_c =1.2\times 10^{38}\,{\rm cm}^{-3}$ ($\sim$\,nuclear density), and a radius $R=10^6\,$cm. In the outskirt of SN, the density and temperature drops as $n(r) = n_c (R/r)^m$ and $T(r) = T_c (R/r)^{m/3}$, with $r>R$ and $m=3-7$. 
The smallest forbidden $\varepsilon$ corresponds to the picture when the $A'$ boson is produced from all over the core, then free streams out of the SN. Similar to the axion case, the process for producing $A'$ is also via bremsstrahlung in proton-proton scattering,
$pp\to pp A'$.
Consider single pion exchange~\cite{Iwamoto:1984ir, Turner:1987by}, the matrix element squared is calculated in~\cite{Dent:2012mx}. 
The cross section is approximately
$\langle\sigma_{pA'}\rangle \approx {6 \varepsilon^2 \alpha m_p T}/({\pi^2 m_\pi^4})$.
As an estimate, the emissivity of $A'$ boson is
$L_{A'} \approx V_c n_c^2 T_c \langle\sigma_{pA'}\rangle = 1.26\times10^{73} \varepsilon^2\,{\rm erg/s}$.
The criterion~\cite{Raffelt:1996wa} for not losing too much energy via $A'$: $L_{A'}<10^{53}\,{\rm erg/s}$, translates into $\varepsilon<0.9\times10^{-10}$.
The largest forbidden $\varepsilon$ corresponds to the trapped picture where $A'$ cannot freely stream out from the core, due to scattering with the medium, $pp A'\to pp$.
The radius of the $A'$ sphere, $r_{A'}$, can be obtained from $\int_{r_{A'}}^\infty n(r) \langle\sigma_{pA'}\rangle(r) dr=2/3$~\cite{Burrows:1986me}. The emission of $A'$ in this case is black body radiation~\cite{thermalization}, $L_{A'} =4\pi r_1^2 \sigma [T(r_{A'})]^4$, with $\sigma=g\pi^2/120$ and the effective degree of freedom (d.o.f.) is $g=3$ for a massive vector boson. The same criterion requires $\varepsilon>(3.1-4.6)\times10^{-7}$ for $m$ between $3-7$. The corresponding $r_{A'}$ ranges between
$2R-10R$. The excluded window obtained in this estimate, $0.9\times10^{-10}<\varepsilon< (3.1-4.6)\times10^{-7}$,
agrees well with those found in Refs.~\cite{Dent:2012mx, Dreiner:2013mua}

Next, let's bring DM into the game. 
Before turning into the core-collapsing SN1987a, the progenitor used to be a star, with about 20 solar mass.
The corresponding radius, temperature, lifetime and escape velocity can be using empirical relations and are summarized in Table~\ref{tab:progenitor}.
\begin{table}[h!]
\begin{tabular}{|c|c|c|c|c|}
\hline
$M/M_\odot$ & $\tau/\tau_\odot$ & $R/R_\odot$  & $T/T_\odot$ & $v_{\rm esc}/v_{\rm esc \odot}$ \\ \hline 
$20$ & $0.75\times10^{-3}$ & $8.1$ & $2.8$ & $1.6$ \\
\hline
\end{tabular} 
\caption{Parameters of the SN 1987a progenitor in the units of those of the Sun. The mass-luminosity and mass-radius relations in Ref.~\cite{progenitor1} are used: $L\sim M^{3.4}$, $R\sim M^{0.7}$. The lifetime is estimated using $\tau \sim M/L$. The temperature of the progenitor is taken from Ref.~\cite{Smith:2006es}. The surface escape velocity satisfies $v_{\rm esc} \sim \sqrt{M/R}$.
}\label{tab:progenitor}
\vspace{-0.3cm}
\end{table}
I further make the assumptions that the progenitor lives in a similar environment as that of the Sun, {\it i.e.}, with similar DM wind velocity $v_{\rm wind}$, local DM number density $n_{\chi}$ and velocity distribution. Therefore, as time goes by, the progenitor will accumulate the DM that ran into it, which satisfies the equation~\cite{Griest:1986yu,Zentner:2009is},
\begin{eqnarray}\label{eq:capture}
\begin{split}
\frac{d N_\chi}{dt} = C_c + C_s N_\chi - A N_\chi^2 \ .
\end{split}
\end{eqnarray}
The first term is the DM capture rate with protons as target, 
$C_c = \sqrt{{3}/{2}} n_\chi \sigma_{\chi p} v_{\rm esc} ({v_{\rm esc}}/{\bar v}) N_p \langle\hat\phi_p \rangle ({{\rm erf}(\eta)}/{\eta})$~\cite{Press:1985ug, Gould:1987ir},
where $v_{\rm esc}$ is the surface escape velocity, and $N_p$ is the total number of protons proportional to stellar mass~\cite{comment2}. 
One can obtain the $C_c$ for the progenitor with the parameters in Table.~\ref{tab:progenitor}, by rescaling from the case of the Sun~\cite{Bertone:2004pz},
\begin{eqnarray}
C_c = 1.6\times 10^{29}\,{\rm s}^{-1} \left( \frac{1\,{\rm GeV}}{M_\chi} \right)^2 \left( \frac{\sigma_{\chi p}}{10^{-39}\,{\rm cm^2}} \right) \ ,
\end{eqnarray}
where the differences in the escape velocity average within the star $\langle\hat\phi_p \rangle$ is also neglected.
For symmetric DM case, the annihilation rate per pair is~\cite{Jungman:1995df},
$A = {(\sigma v)_{\rm anni}}/{V_{\rm eff}} = (\sigma v)_{\rm anni} [ ({M_\chi \rho})/({3 M_{\rm pl}^2 T})]^{3/2}$.
Under the approximation, $\rho \propto M/R^3$,
\begin{equation}
A= 8.2\times10^{-60}\,{\rm s}^{-1} \left( \frac{(\sigma v)_{\rm anni}}{3\times10^{-26}\,{\rm cm^3/s}} \right) \left( \frac{M_\chi}{1\,{\rm GeV}} \right)^{3/2}.
\end{equation}
From~\cite{Zentner:2009is}, the DM self capture rate per capita is,
\begin{equation}
C_s=4.3\times 10^{-15}\,{\rm s}^{-1} \left( \frac{1\,{\rm GeV}}{M_\chi} \right)^2 \left( \frac{\sigma_{\chi \chi}}{10^{-24}\,{\rm cm^2}} \right) \ , 
\end{equation}
assuming similar $\langle\hat\phi_\chi \rangle=5.1$ for the progenitor and the Sun. 

Because the progenitor has much larger mass than the Sun and thus larger escape velocity from the core region, the evaporation effect is neglected for $M_\chi\gtrsim1\,$GeV.

\begin{figure}[t!]
\includegraphics[width=1.0\columnwidth]{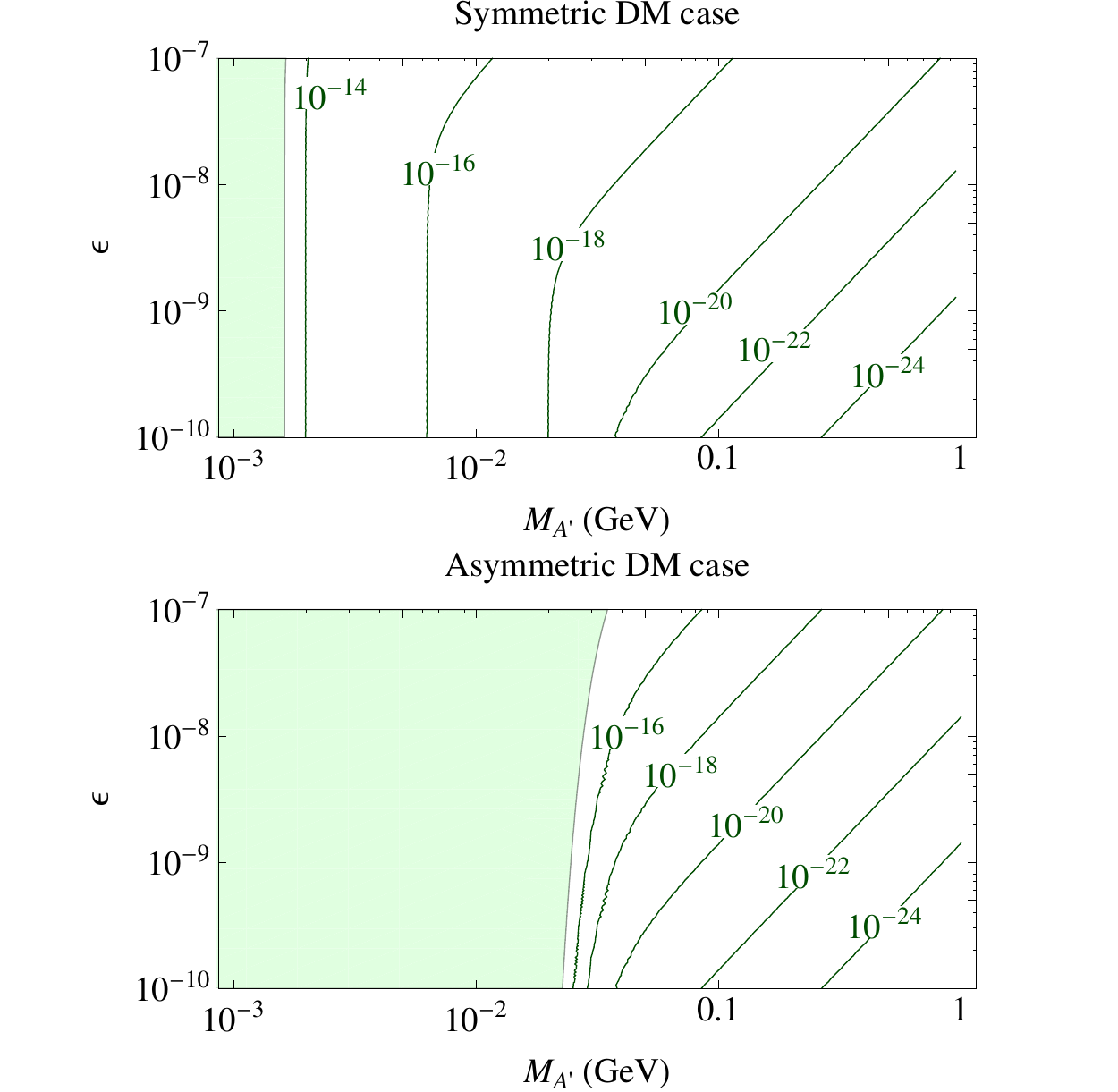}
\vspace{-0.2cm}
\caption{The ratio of captured DM number to proton number inside the progenitor (or SN), with fixed $M_\chi=1\,$GeV and $\alpha'=0.03$. 
In the green shaded regions, this ratio saturates to the dark disk limit.
\vspace{-0.3cm}}\label{fig:NX}
\end{figure}

The general solution to Eq.~(\ref{eq:capture}) takes the analytic form~\cite{Zentner:2009is}
\begin{eqnarray}
\begin{split}
N_\chi(\tau) = \frac{C_c \tanh (\zeta^{-1}\tau)}{\zeta^{-1} - \frac{1}{2}C_s \tanh (\zeta^{-1}\tau)} \ ,
\end{split}
\end{eqnarray}
where $\zeta^{-1} = \sqrt{C_s^2/4 + A C_c}$. When $C_s>\sqrt{A C_c}$, DM self capture rather than proton capture dictates the final number of captured DM.
In the case of asymmetric dark matter, $A\to0$, the captured number exponentially grows with time, $N_\chi(\tau) = (C_c/C_s) [\exp(C_s \tau)-1]$. 
A general upper bound on the growth rate exists when approaching the dark-disk limit~\cite{Frandsen:2010yj}: $d N_{\chi}(t)/d t \leq n_\chi \pi R^2 v_{\rm wind}$. The righthand side equals $6.6\times10^{30}\,{\rm s}^{-1} (1\,{\rm GeV}/M_\chi)$, by using $R_\odot=7\times10^{5}\,$km, $n_\chi = 0.3 \times10^{15} \,{\rm km}^{-3} (1\,{\rm GeV}/M_\chi)$~\cite{Salucci:2010qr}, $v_{\rm wind}=220\,{\rm km/s}$, and a rescaling with Table~\ref{tab:progenitor}.

Given the hidden U(1) theory, Eq.~(\ref{theory}), one can calculate the relevant cross sections for DM scattering and annihilation.
For $M_\chi>m_{A'}$, $\chi\bar\chi$ can annihilate into a pair of $A'$ gauge bosons, with a cross section
$(\sigma v)_{\rm anni} = ({\pi \alpha'^2}/{M_\chi^2}) \sqrt{1- {m_{A'}^2}/{M_\chi^2}}$,
where $\alpha' = e'^2/(4\pi)$ is the hidden fine structure constant. 
In direct detection, the DM elastically scatters with proton via $A'$ exchange and the cross section is spin independent,
$\sigma_{\chi p} ={16\pi \varepsilon^2 \alpha \alpha' \mu_p^2}/{m_{A'}^4}$,
where $\mu_p = m_p M_\chi/(m_p + M_\chi)$ and $m_p$ is the proton mass.
The model considered also features dark matter self interaction, mediated by $A'$ exchange. 
The Born level cross section is,
$\sigma_{\chi \chi} = {4\pi \alpha'^2 M_\chi^2}/{m_{A'}^4}$.
For large enough $\alpha'$, non-perturbative and many-body effects may be important~\cite{Tulin:2012wi}.
I will neglect them in this work, to be on equal footing with the single pion exchange treatment
in the $A'$-bremsstrahlung from proton scatterings.

Fig.~\ref{fig:NX} plots the number of DM captured by the progenitor before it collapses, in units of proton number. The two cases of symmetric and asymmetric DM are shown, with fixed $M_\chi=1\,$GeV and $\alpha'=0.03$.
For asymmetric DM, due to the absence of annihilation, when $m_{A'}$ reduces to a few tens of MeV, the self capture effect is so strong that the dark disk limit is quickly saturated, which dictates
$N_\chi/N_p \lesssim n_\chi \pi R^2 v_{\rm wind} \tau/N_p = 6.5\times10^{-14}$.
This ratio is much lower than that can be achieved in the Sun~\cite{Frandsen:2010yj}, largely because of the shorter lifetime for more massive star.
In symmetric DM case, the captured dark matter number is relatively smaller due to annihilation~\cite{comment}.
For $M_\chi\gtrsim 1\,$GeV, the thermally produced DM number density is found to have a negligible effect.

When the time comes, gravity forces electrons and protons to turn into neutrons and neutrinos, which causes the iron core to collapse into a young neutron star. The captured DM are also likely to resettle around the core. 
The thermalization radius of DM, $r_{\rm th} = \sqrt{9 T_c/(8\pi G_N M_\chi \rho_c)} \sim 10\,$km for $M_\chi \sim1\,$GeV, is roughly the same as the size of the SN core.
Here I make a simplified ``comoving'' assumption such that the DM distribution follows the same shape as protons, only rescaled by the ratio of total particle number determined above, {\it i.e.}, $n_\chi(r)/n_p(r) = N_\chi/N_p$. 

Typically, this amount of DM is too tiny to affect the production of $A'$, {\it i.e.}, compared to the dominant production channel $pp\to pp A'$, the $\chi p\to\chi p A'$ rate is suppressed by a factor $N_\chi/N_p$, while the $\chi\chi\leftrightarrow\chi\chi A'$ process is further down by $(N_\chi/N_p)^2(1/\varepsilon)^2\ll1$, for the values of $\varepsilon$ of interest.

However, it is much easier for DM to play an important in deflecting/trapping the $A'$ that have been produced. 
The relevant process is the hidden sector analog of the low-energy Thomson scattering, 
$A' \chi \to A' \chi$, whose cross section is
\begin{eqnarray}
\sigma_{\chi A'} = \frac{8\pi}{3} \frac{\alpha'^2}{M_\chi^2} = 2.9\times10^{-30}\,{\rm cm}^2 \left( \frac{\alpha'}{0.03} \right)^2 \left( \frac{1\,{\rm GeV}}{M_\chi} \right)^2 \ . \hspace{-0.35cm} \nonumber \\
\end{eqnarray}
Comparing with the conventional trapping process $pp A'\to p p$, although the Thomson scattering rate here is suppressed by the target number density $N_\chi/N_p$, 
its cross section has a relative enhancement factor $[\alpha'/(\varepsilon \alpha)]^2$.
Because the SN cooling constraint is sensitive to the regime $10^{-10} < \varepsilon < 10^{-7}$, the relative enhancement factor can be large enough to win over the $N_\chi/N_p$ suppression.

Sufficiently large Thomson scattering increases the opacity to $A'$ and can already create a (pseudo) $A'$ sphere, inside which $A'$ cannot escape. 
In general, the sphere radius can be found with $\int_{r_{A'}}^\infty [n_\chi(r) \sigma_{\chi A'} + n_p(r) \sigma_{p A'}(r)] dr=2/3$. It is useful to define a quantity 
\begin{eqnarray}
\begin{split}
\sigma_0 = \frac{2}{3 (n_\chi)_c R} = 5.6\times 10^{-30}\,{\rm cm}^2 \left( \frac{10^{-15}}{N_\chi/N_p} \right) \ .
\end{split}
\end{eqnarray}
For simplicity, I make the approximation by neglecting the $r$ dependence in $\sigma_{p A'}$ hereafter. There are three possible cooling phases:

{\bf i)} At very low DM density, $\sigma_{\chi A'} + (n_p/n_\chi) \sigma_{p A'} < (m-1) \sigma_0/m$, there exists no (pseudo) $A'$ sphere. The usual SN bound applies.

{\bf ii)} At intermediate DM density, when $(m-1) \sigma_0/m <\sigma_{\chi A'} + (n_p/n_\chi) \sigma_{p A'} < (m-1) \sigma_0$, the (pseudo) $A'$ sphere is within the core, $0<r_{A'}<R$.
The emissivity of $A'$ is proportional to the part of core volume outside the sphere,
\begin{eqnarray}
\begin{split}
L_{A'} \approx \left(V_c - \frac{4}{3}\pi r_{A'}^3 \right) n_c^2 T_c \langle\sigma_{pA'}\rangle \ ,
\end{split}
\end{eqnarray}
where $V_c=4\pi R^3/3$, and $r_{A'} = [m/(m-1) - \sigma_0/(\sigma_{\chi A'} + (n_p/n_\chi) \sigma_{p A'})] R$.
The constraint on the smallest forbidden $\varepsilon$ will be relaxed by a factor of
$\sqrt{{R^3}/({R^3 - r_{A'}^3})}$.

{\bf iii)} At sufficiently high DM density, $\sigma_{\chi A'} + (n_p/n_\chi) \sigma_{p A'} > (m-1) \sigma_0$, the (pseudo) $A'$ sphere is located in the outskirt of the SM, $r_{A'}>R$.
The emissivity to cool the core becomes
\begin{eqnarray}\label{11}
\begin{split}
L_{A'} =4\pi r_{A'}^2 \sigma & T_c^4 \left( \frac{R}{r_{A'}} \right)^{{4m}/{3}} \left[ \frac{n_p \sigma_{p A'}}{n_\chi \sigma_{p A'} + n_p\sigma_{\chi A'}} \right],
\end{split}
\end{eqnarray}
with $r_{A'} =[(\sigma_{\chi A'} + (n_p/n_\chi) \sigma_{p A'})/ ((m-1)\sigma_0)]^{1/(m-1)} R$.
The last factor in Eq.~(\ref{11}) reflects the fact that when $\varepsilon$ (and thus $\sigma_{p A'}$) is tiny, 
$A'$ is not chemically thermalized inside the pseudo sphere, the effective d.o.f. is suppressed.
The emissivity is suppressed compared to blackbody radiation in thermalized case.

\begin{figure}
\includegraphics[width=1.0\columnwidth]{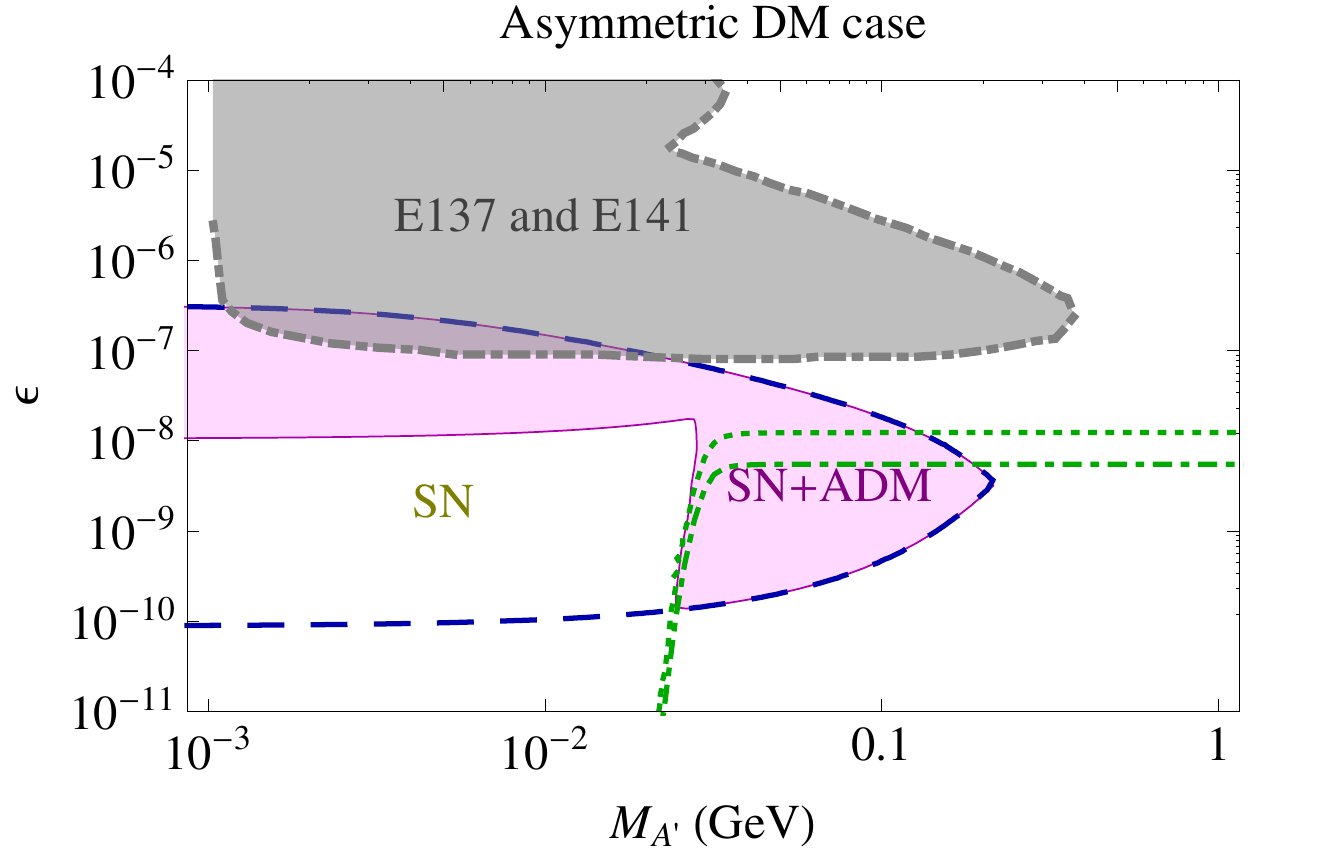}
\vspace{-0.4cm}
\caption{SN cooling constraint on the $\varepsilon-m_{A'}$ plane, with the same set of parameters as Fig.~\ref{fig:NX}. The region enclosed by blue long dashed curve is excluded without consider DM. 
The excluded region shrinks to the magenta shaded region when additional opacity due to DM is taken into account.
The green dot-dashed curve corresponds to the (pseudo) $A'$ sphere just appearing from the origin, and the dotted curve corresponds to the sphere crossing the edge of SN core.
The E137 and E141 experimental exclusion is shown in the gray region.
}\label{fig:comb}
\vspace{-0.3cm}
\end{figure}

In Fig.~\ref{fig:comb}, I plot the constraints on the $\varepsilon-m_{A'}$ parameter space, with mass at 1\,GeV and $\alpha'$ equal to 0.03. 
For $m_{A'}$ larger than the core temperature, a Boltzmann suppression factor is multiplied to the emissivity. 
The decay length of $A'\to e^+e^-$ is also required to be longer than the radius $R$ when the cooling constraint applies.
As discussed, in the absence of DM, SN cooling by emitting the hidden gauge boson $A'$ are sensitive to $\varepsilon$ values between $10^{-10} - 10^{-7}$.
In this case, the SN exclusion region is enclosed by the dashed yellow curve.
In contrast, when the DM is taken into account, the SN cooling exclusion region shrinks to the magenta regions. 
The sudden change in the exclusion near $m_{A'}\approx 30\,$MeV is due to the drastic change in the captured DM number density.
There, the DM is abundant enough to efficiently increase the opacity to $A'$ and lower its emissivity.

The impact of DM's presence can be important. 
From Fig.~\ref{fig:comb}, it opens a window which increases the upper bound on $\varepsilon$ by 2 orders of magnitude, 
allowing it to be as large as $\sim10^{-8}$. 
This happens at $m_{A'}\lesssim 20-30\,$MeV for the asymmetric DM case (for symmetric DM, the window opens at $m_{A'}\lesssim 2\,$MeV).
The effect would get stronger if the assumptions made on the DM density and DM-hidden boson coupling are relaxed.

Reopening the $\varepsilon$ window could be interesting for theoretical model building and motivate new experimental searches.

To conclude, I discussed a mechanism where the dynamics of SN cooling via emitting exotic light particles can be strongly affected by the existence of DM.
I discuss a simple hidden sector model containing a light hidden gauge boson with kinetic mixing with the photon, and DM charged under it.  
Inside SN, a smog of DM shortens the free streaming length of the hidden gauge boson, thus increases the opacity to it.
It is reasonable to expect that the progenitor of SN have been capturing dark matter throughout the whole lifetime, for the above effect to take place.
I have focused on GeV scale DM, and in particular the case of asymmetric DM, 
and showed the constraints inferred from the observation of SN1987a can be relax by as large as two orders of magnitude. 
It can be worthwhile to consider a wider range of DM masses.
As a sketch, for heavier DM, this effect is weaker because both its local number density and the Thomson scattering rate are suppressed, meanwhile, the direct detection limit from say CDMSlite could become relevant~\cite{Agnese:2013jaa};
For lighter DM, the evaporation effect during capture is not negligible, but the thermal production of DM could catch up and dominate. 
In the latter case, the emission of DM itself may also be an important process.
Better knowledge of the astrophysical parameters related to the SN and its progenitor and more accurate calculation of the energy transfer rates
would also help to reach a more quantitative and complete picture.

\medskip
\noindent{\it Acknowledgements.} \ \ I would like to thank Haipeng An, Clifford Cheung, Rabi Mohapatra, Goran Senjanovi\'c, Mark Wise and Haibo Yu for useful discussions. This work is supported by the Gordon and Betty Moore Foundation through Grant \#776 to the Caltech Moore Center for Theoretical Cosmology and Physics, and by the DOE Grant DE-FG02-92ER40701.


\begin{thebibliography}{99}
\bibitem{thermalization} 
  The cost of free energy for fully thermalizing the $A'$ number density inside the core of SN is roughly $V_c n_{A'}^{\rm eq} T_c \simeq 10^{50}\,{\rm erg}$, too small to affect the cooling rate with neutrinos $10^{53}\,{\rm erg}/s$, which lasts for several seconds.
\bibitem{comment2}
Here only the DM scattering with proton which makes up hydrogen is considered. 
In the later stage of the star when heavy elements are abundantly synthesized, there could be coherent enhancement in the DM-nucleus scattering cross section. 
This effect is not considered because such stage lasts for a much shorter time scale than hydrogen burning.
\bibitem{comment} 
This result depends on the assumption that the DM density near the SN is similar to that of the Sun.
More DM can be captured if the local density is higher, or if the DM self interaction is stronger.
If it is the case, even the symmetric DM case could strongly impact the SN cooling, as discussed below.
\end{thebibliography}
\end{document}